# The influence of persona and conversational task on social interactions with a LLM-controlled embodied conversational agent

Leon O. H. Kroczek[1*], Alexander May[1], Selina Hettenkofer[1], Andreas Ruider[1], Bernd Ludwig[2], & Andreas Mühlberger[1]

[1]*Department of Psychology, Clinical Psychology and Psychotherapy, University of Regensburg*
[2]*Chair of Information Science, University of Regensburg*

**Abstract**
Large Language Models (LLMs) have demonstrated remarkable capabilities in conversational tasks. Embodying an LLM as a virtual human allows users to engage in face-to-face social interactions in Virtual Reality. However, the influence of person- and task-related factors in social interactions with LLM-controlled agents remains unclear. In this study, forty-six participants interacted with a virtual agent whose persona was manipulated as "extravert" or "introvert" in three different conversational tasks (small talk, knowledge test, convincing). Social-evaluation, emotional experience, and realism were assessed using ratings. Interactive engagement was measured by quantifying participants' words and conversational turns. Finally, we measured participants' willingness to ask the agent for help during the knowledge test. Our findings show that the extraverted agent was more positively evaluated, elicited a more pleasant experience and greater engagement, and was assessed as more realistic compared to the introverted agent. Whereas persona did not affect the tendency to ask for help, participants were generally more confident in the answer when they had help of the LLM. Variation of personality traits of LLM-controlled embodied virtual agents, therefore, affects social-emotional processing and behavior in virtual interactions. Embodied virtual agents allow the presentation of naturalistic social encounters in a virtual environment.

## 1. Introduction

The rise of large language models (LLMs) has revolutionized human-computer interaction, enabling more sophisticated and intuitive exchanges between people and machines. LLMs can be instated as conversational agents (e.g. ChatGPT), allowing users to engage in meaningful language-based interactions. While earlier versions of conversational agents were based on simple pattern-matching algorithms that generated pre-defined output (Ramesh et al., 2017), LLMs use probabilistic generative processes based on training data to generate responses (Generative Pre-Trained Transformer, GPT; Yenduri et al., 2024). Furthermore, because LLMs can consider the contextual information of an input, they allow the generation of individualized responses that match the user's intention. Because of these properties LLM controlled conversational agents are increasingly being used in chat-based interactions for example in customer service and healthcare (Rivas & Zhao, 2023; Thirunavukarasu et al., 2023). However, while chat-based interactions are useful in many tasks, face-to-face conversations may be advantageous as they allow the simulation of ecological valid human-to-human interactions and the presentation of additional information via non-verbal cues (e.g., gesture, gaze, facial expressions). Importantly, face-to-face social interactions are better suited for establishing a deeper social connection between interactive partners than text based interactions (Sacco & Ismail, 2014). This has inspired the development of embodied conversational agents, in which computer-based dialogue systems are combined with animated virtual agents (Huang, 2018). Presenting embodied conversational agents in Virtual Reality (VR) as 3D, life-like virtual humans enable users to engage in multimodal, face-to-face conversation with an interactive partner in front of them. As with chatbots, recent attempts have been made to increase the naturalism of conversations by incorporating LLMs as embodied conversational agents (Hasan et al., 2023; Lim et al., 2024). The high degree of flexibility and individualization of LLM controlled conversations makes them superior to scripted speech stimuli implemented in virtual scenarios that can only follow a pre-defined conversational path. Another benefit of LLMs is that they are pretrained for standard conversation in natural language. Without the need to collect huge, but specialized data corpora for training dialogue models, LLMs can be integrated immediately in virtual agents. This allows use cases in research focused on social interactions as well as applications, for example conversations with virtual patients can be used in medical education (Graf et al., 2024) and extended conversations could be implemented in virtual exposure therapy .However, while such set-ups can have a profound impact on human experience and behavior, little is known about the underlying mechanisms and contextual influences that drive social interactions with LLM controlled embodied virtual agents.

In real-life social interactions people are fast to form social evaluations of their interactive partners (Bar et al., 2006; Satchell, 2019) and similar effects have been found for interactions with virtual agents (Guadagno et al., 2011). According to the computers as social actors framework these results can be explained by the automatic and unconscious response to social cues regardless of whether these cues are produced by a computer or a human (Nass & Moon, 2000). In addition, it has been argued that behavioral realism, i.e. the degree to which an agent's social cues resemble human ones, plays a crucial role in this regard (Von Der Pütten et al., 2010) with more behavioral realism resulting in more social behavior towards the agent. Regarding human-AI interactions, previous findings support the computer as social actors framework for interactions with chatbots, e.g. by showing that an agent's empathic expressions lead to a more favorable social evaluation than neutral expressions (Liu & Sundar, 2018).

Furthermore, there is evidence of the impact of personality traits and communicative style of conversational agents and robots. For instance, in text-based communication an agent's assertive and serious personality compared to a warm and cheerful personality increased the willingness to confide in the agent during a job interview (Zhou et al., 2019). With regard to the specific personality traits of conversational agents it was shown that chatbots scoring high on extraversion resulted in higher social presence and increased communicative satisfaction compared to less extraverted chatbots (Ahmad et al., 2021). Overall, the communicative style and personality traits of a conversational agent determine how users experience the interaction with an agent. However, this has been mostly tested in text-based communication or with pre-determined responses, and it remains an open question whether similar results can be observed for LLM-controlled agents who are presented face-to-face as an embodied interactive partner in Virtual Reality.

Conversational agents built on LLMs can be used for a wide variety of tasks. They can be used to answer specific questions about world knowledge, provide detailed instructions about how to perform a particular task, give everyday advice, or serve as social companions in small-talk or even deeper conversations (Skjuve et al., 2023). Importantly, the nature of a task or context may influence how users interact with a conversational agent. For instance, receiving help from a conversational agent may generally lead to a more pleasant experience. Furthermore, tasks may interact with personality traits or communicative style, in a sense that particular traits might be positively evaluated in some tasks but not in others. Only few studies have investigated this relationship. Roy et al. (2021) found that extraverted agents were more favorably rated than introverted agents in tasks where the goal was to provide information, whereas there were no clear preferences in tasks where the goal was to complete an assignment. Another study demonstrated that users' competency levels may also play a role in this regard, as high-competency users rated a social-oriented interactive style as more useful, while low-competency users rated a task-oriented style as more useful when interacting with a conversational agent (Chattaraman et al., 2019). Taken together, tasks may play an important role in how user's form social evaluations about conversational agents; however, studies that investigate LLM-controlled embodied virtual agents across different tasks are missing.

The current study was designed to test how persona descriptions of an LLM-controlled virtual agent and conversational tasks would affect experience and behavior in face-to-face social interactions. Participants engaged in face-to-face social interactions with an LLM-controlled agent in Virtual Reality. The persona of the LLM was manipulated to be either "extraverted" or "introverted". Three conversational tasks were conducted: Small Talk, Knowledge Test, and Convincing. Tasks were chosen to reflect a social conversation focused on self-disclosure (small talk), task-oriented conversation (knowledge test), and social conversation focused on argumentation (convincing). Ratings were assessed to measure the social evaluation of the agent (sympathy and closeness), emotional experience (valence and arousal), and the experience of the virtual interaction (realism and social presence). In addition, the number of words and turns as well as the number of times participants asked the agent for help in the knowledge test were measured to quantify interactive engagement.

In a set of preregistered hypotheses we expected (1) that the small talk conversation would be experienced more pleasant than the knowledge test and the argumentation during the convincing task due to increased intimacy during self- and other-disclosure (Park et al., 2011). (2) Furthermore, we expected that the evaluation of the agents persona would differ as a function of conversational task with respect to sympathy, valence, closeness, and realism, with higher ratings for the extraverted compared to the introverted agent. This difference was expected to be more pronounced in the small talk than in the knowledge and convincing task. (3) In this line, we also expected an interaction effect between persona and conversational tasks on interactive engagement, i.e. the number of words spoken by the participant during a conversation: more words were expected in the extraverted compared to introverted condition. This difference was expected to be greater in the small talk task than in the knowledge and convincing tasks. (4) Finally, it was hypothesized that participants would ask the extraverted agent more frequently for help than the introverted agent, as the barrier to ask for help was expected be lower when social interactions were more pleasant due to the extraverted agent being more talkative and social (Ashfaq et al., 2020).

## 2. Methods
### 2.1. Participants

Forty-six healthy volunteers were recruited at Regensburg University (mean age = 21.24 years, SD = 2.59, 36 females). Participants were randomly assigned into two groups of 23 participants with groups being matched for gender. A sensitivity analysis conducted using G*Power (Faul et al., 2009) revealed that this sample size allowed the detection of a medium to large effect size of $\eta_p^2 = .10$ with a power of 0.8 for a mixed effect ANOVA design. Participants did not report mental or neurological disorders and had normal or corrected-to-normal vision. All participants gave written informed consent. The study was conducted in line with the Declaration of Helsinki and was approved by the ethics committee of Regensburg University (24-3669-101). Participants received credit points for compensation.

### 2.2. Experimental-Design

The study used a mixed design with the agent's *persona* (extraverted vs. introverted) as a between-subjects factor and *conversational task* (small talk vs. knowledge test vs. convincing) as a within-subject factor. During the knowledge test an additional manipulation was introduced by alternating between *easy* and *difficult* questions that had to be answered by the participants. Sympathy, Valence, Arousal, Closeness, and Naturalism were measured via self-report after each conversational task and number of words and turns during each task were assessed as behavioral measures. In the knowledge test, the percentage of questions for which the participants asked the agent for help was assessed as an additional metric. Finally, the social presence subscale of the multimodal presence scale (MPS, Makransky et al., 2017) was measured for each conversational task.

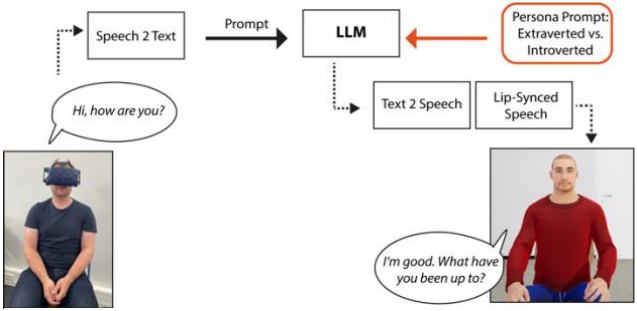

*Figure 1: Schematic overview of systems components.*

### 2.3. Apparatus
#### 2.3.1. Technical Set-Up

All system components are shown in the schematic overview in Figure 1. Virtual Reality was presented via a head-mounted display (Vive Pro, HTC, Taoyuan, Taiwan). The virtual environment was rendered and controlled using Unreal Engine 5.2 (Epic Games Inc., Raleigh, USA). Participants were placed in a virtual room at a table face-to-face with a male virtual agent (MetaHuman. Epic Games Inc.). The same agent was used in both between-subject personality conditions and the distance between participant and agent was set to 1.8 meter. Eye gaze of the virtual agent was controlled so that the agent was always looking towards the participant. Participant speech was sampled with an external USB microphone (t-bone SC 420, Thomann, Germany) and converted to text using the wav2vec model (Baevski et al., 2020). The text input was then forwarded as a prompt to a text generation model ("SauerkrautLM-HerO", https://huggingface.co/VAGOsolutions/SauerkrautLM-7b-HerO) which was run on a separate computer using the text generation inference toolkit with streaming enabled. Importantly, deploying all components (including the LLM) on local servers allowed us to maintain full control over all data without the need to rely on third-party services and in accordance to data protection policies. Across all experimental sessions, the average time of the LLM to generate a response was 1.29 sec (SD = 0.26, range = 0.68 – 4.11 sec).

Two different personas were implemented as context prompts to the text generation model, i.e. prompts that preceded the user input, depending on the persona manipulation "extravert" or "introvert". The personas included details about the communicative style and gave examples on how the agent would answer particular questions (see Table 1, full text in supplementary material). The descriptions varied in how persona were described as being sociable, talkative, and enjoying engaging with other people which all relates to the dimension of extraversion in the five-factor theory of personality (McCrae & John, 1992). Each persona also included the same information about how the LLM had to answer specific questions about preferences. This information was included to ensure that no differences in sympathy based on preferences (e.g. liking dogs better than cats) could arise between personas. Additional parameters, such as the maximum number of generated sentences and the temperature parameter, were manipulated between the personas (Table 1). The text output of the LLM was then converted into speech (male voice) using a speech2text model (Silero, https://github.com/snakers4/silero-models). The same voice was used for all personality conditions. Finally, the resulting audio was streamed to the Audio2Face application (Nvidia Omniverse, https://build.nvidia.com/nvidia/audio2face) where live lip movements were animated based on the "Mark" set-up. Animations and audio were then played by the virtual agent using the LiveLink plugin in the Unreal Engine.

#### 2.3.2. Ratings and Questionnaires

After each conversational task, participants were asked to rate their experience using a visual analog scale (values 0 - 100). Ratings included dimensions to describe emotional experience (valence and arousal), as well as social evaluation of the virtual agent (sympathy and closeness). These ratings included sympathy ("*How sympathetic was the virtual person?*", 0 = very unsympathetic, 100 = very sympathetic), valence ("*How unpleasant or pleasant did you feel during the interaction?*", 0 = very unpleasant, 100 = very pleasant), arousal ("How high was your arousal during the interaction"), closeness ("*How close/connected did you feel to the virtual person?*", 0 = very distanced/unconnected, 100 = very close/connected), realism ("How unnatural or natural did you experience the interaction?" 0 = very unnatural, 100 = very natural). Because the "knowledge test" conversational task required participants to answer questions about general world knowledge, two additional ratings were included that asked for participants' confidence in the correctness of their answer for questions answered with or without help of the virtual agent ("*How confident are you that your answers were correct for questions that you answer on your own/for questions that you answered with help of the virtual person?*", 0 = very unconfident, 100 = very confident).

*Table 1: Short persona description and LLM parameters. For full persona descriptions see supplementary material.*

| Persona | Prompt | Temperature | Maximum Sentences |
|---|---|---|---|
| Extravert | Your name is Moritz. You have a very extroverted character. This means you are very sociable, enjoy being around people, are active, and talkative. You engage a lot with your conversation partner and ask interested follow-up questions. […] | 0.4 | 3 |
| Introvert | Your name is Moritz. You are a very introverted character. This means you enjoy being alone, are independent, and very reserved in social interactions. […] | 0.2 | 2 |

Social anxiety symptoms were assessed using the Social Phobia Inventory (SPIN ,Connor et al., 2000). In addition, the social presence subscale of the Multimodal Presence Scale (MPS; Makransky et al., 2017) was used to measure the experience of social presence in each conversational task. The subscale averages over five items that have to be answered on a 5-point Likert scale. Furthermore, a German translation of the General Attitudes towards Artificial Intelligence Scale was used (GAAIS, Schepman & Rodway, 2020). The questionnaire included 16 positive items (e.g. opportunities, benefits, and positive emotions related to AI) and 16 negative items (concerns and negative emotions related to AI). Items were answered on a 5-point Likert scale, and average scores were calculated for each subscale. Finally, personality traits were measured using the 30 item short version of the NEO five-factor-inventory (NEO-FFI-30, Körner et al., 2008). Data from questionnaires are accessible in the open data but were not analyzed in the present study.

### 2.4. Procedure

After providing written informed consent, participants were asked to fill in questionnaires about demographic information (age, sex, and occupation) and social anxiety symptoms, participants then sat on a chair and put-on the HMD where the virtual room was presented. Participants position in the virtual room was adjusted so that they were sitting at the virtual table, directly facing the virtual agent. The virtual agent was presented from the beginning, but the LLM was not activated before the conversational task, i.e. there was no speech by the agent before the first conversational task.

Next, the first conversational task was initiated by the experimenter. The small talk conversation was always implemented as the first task in order to simulate a natural interaction in which persons first get to know each other. Instructions about the task were displayed on a virtual screen on the table and participants used the motion controller to forward through instructions. Participants were informed that they would be prompted to verbally ask the virtual agent small talk questions displayed on the virtual screen and that they could use the motion controller to forward to the next question. Participants were also specifically instructed that they were free to follow-up on anything the virtual agent said. The conversational task was then conducted for a total of eight minutes, during which the LLM was set active so that responses were generated for any input that was produced by the participants. The question which participants had to ask about related to general information about the agent (e.g. "*Do you have sisters or brothers?*") and were taken in part from the small talk questions described in Aron et al. (1997). After 8 minutes, participants took off the HMD, answered rating questions, and filled in the social presence subscale of the MPS.

Afterwards, participants put on the HMD again and the conversational task "knowledge test" was started in which participant had to answer short questions on world knowledge (e.g. "*What is the highest building in the world?*"). Before the task, participants were informed about the task, and it was highlighted that they were free to ask the virtual agent for help. In contrast to the previous task, participants had to activate agent's response mode (where the LLM generated responses for input) by pressing a button. The response mode was automatically deactivated after the agent/LLM had responded. Questions alternated between questions that were relatively easy and relatively hard to answer (based on the experimenter's evaluation). Participants had to give their answer verbally and could then forward to the next question using the motion controller. The knowledge test lasted for 8 minutes and was followed by ratings and the social presence subscale of the MPS outside Virtual Reality.

"Convincing" was conducted as a last conversational task where participants had to ask the virtual agent about preferences regarding two options (e.g. "*What do you prefer: cats or dogs?*"). Following the agent's response, participants then had to convince the agent of the alternative that had not been chosen by the agent (also regardless of their own preferences). In addition, participants were instructed that they were free in how many attempts they would try to convince the virtual agent. Preference questions were displayed on the virtual screen and participants could forward to the next question using the motion controller. The conversational task had a duration of 8 minutes and was followed by a final set of ratings and the social presence questionnaire.

After the last conversational task participants were asked to fill in questionnaires regarding their general attitudes towards AI (GAAI), and personality (NEO-FFI 30). The experiment had a total duration of about 1.5 hours.

### 2.5. Pre-test: Validation of LLM persona

To ensure that the personality manipulation of the LLM resulted in the intended changes a personality test (NEO-FFI-30) was conducted for each persona prompt. To account for variability in model output, the questionnaire was administered 100 times with both personas. Reliability was calculated across repetitions using Cronbach's alpha. Consistency was estimated at .87 for the extraverted persona and at .92 for the introverted persona demonstrating good reliability for both personas. Personas were compared for each NEO-FFI subscale. Results revealed that, across repetitions, the extraverted persona scored significantly higher than the introverted on the extraversion subscale (extraverted: $M = 12.81$, $SD = 4.21$, introverted: $M = 7.87$, $SD = 3.09$), $t(181.78) = 9.64$, $p < .001$, $d = 1.36$, while there were no differences on the neuroticism (extraverted: $M = 9.38$, $SD = 3.50$, introverted: $M = 9.39$, $SD = 2.96$), $t(192.85) = -0.02$, $p = .983$, $d < 0.01$, consciousness (extraverted: $M = 9.83$, $SD = 3.51$, introverted: $M = 9.30$, $SD = 3.04$), $t(194.02) = 1.14$, $p = .510$, $d = 0.16$, and openness subscale (extraverted: $M = 8.46$, $SD = 3.81$, introverted: $M = 7.51$, $SD = 3.26$), $t(193.39) = 1.90$, $p = .178$, $d = 0.27$. On the agreeableness subscale the introverted persona scored higher than the extraverted persona (extraverted: $M = 5.44$, $SD = 3.09$, introverted: $M = 7.70$, $SD = 3.01$), $t(197.88) = -5.24$, $p < .001$, $d = 0.74$. Overall, the largest effect of persona was observed with respect to extraversion.

## 2.6. Data Preprocessing and Statistical Analysis

During the experiment participants' speech was transcribed using the speech-to-text model (see above) and written to a logfile. These data were analyzed by counting the number of words and turns (e.g. an uninterrupted segment of speech of the participant with no interleaved response from the virtual agent) for each question during a conversational task. Note that to characterize interactive engagement, words and turns were analyzed per question rather than as a total number. This was necessary because otherwise simply asking the displayed question and then switching to the next question without any true interactive engagement would have increased the total number of words or turns (and this behavior would have been more likely to occur for the introverted persona where responses were short). One participant was excluded from the analysis because only one question was asked.

Statistical analysis were conducted in the R environment (R Core Team, 2016). Data for each dependent variable were analyzed using a mixed effect ANOVA with persona as between-subject and conversation task as a within-subject factor. Sphericity violations were corrected using Greenhaus-Geisser correction. In case of significant interaction effects post-hoc t-test were conducted using Holm method to correct for multiple comparisons. Alpha level was set at .05.

## 2.7. Open Science Statement

Study procedures and hypotheses were preregistered (https://osf.io/tukqh/?view_only=977721578f4d42a5bf9417d6d08bbe46). Ratings, questionnaires, and secondary data on conversations (word number, turns) as well as analysis scripts are publicly available in an online repository (https://osf.io/ws7jf/?view_only=7bb7b39b401a444d8ec59bcccb4d2e93). Note that transcriptions of conversations may contain personal information and are, therefore, not publicly available.

## 3. Results
### 3.1. Manipulation Check

The number of words generated by the LLM per turn was compared between personas as a manipulation check. The extraverted persona generated significantly more words (M = 24.19, SD = 75.62) per turn than the introverted persona (M = 15.35, SD = 29.80), $t(41.90) = 11.08$, $p < .001$, $d = 3.30$ and was thus significantly more talkative. These data demonstrate that the personality manipulation via persona prompts was successful and resulted in the intended effects (see also validation of LLM persona above).

### 3.2. Social Evaluation of Virtual Agent

We investigated how interacting with different LLM personas during different conversational tasks affected participants' social evaluation of the virtual agents in terms of sympathy and closeness (Figure 2).

Analysis of sympathy of the virtual agent revealed a main effect of *persona*, $F(1,44) = 21.57$, $p < .001$, $\eta_p^2 = .33$, a main effect of *conversational task*, $F(2, 88) = 9.82$, $p < .001$, $\eta_p^2 = .18$, but no interaction effect between *persona* and *conversational task*, $F(2, 88) = 2.97$, $p = .056$, $\eta_p^2 = .06$. Sympathy was higher for extraverted (M = 58.20, SD = 12.12) than for introverted persona (M = 41.19, SD = 12.71). Sympathy was rated higher for the knowledge test task (M = 56.70, SD = 12.37) than for small talk (M = 45.91, SD = 21.31), $t(45) = 3.77$, $p = .001$, and convincing task (M = 46.47, SD = 46.48), $t(45) = 3.80$, $p = .001$, but there was no significant difference between small talk and convincing, $t(45) = -0.20$, $p = .843$.

Analysis of participants' feelings of closeness towards the virtual agent revealed a main effect of *conversational task* $F(2, 88) = 3.69$, $p = .029$, $\eta_p^2 = .08$, but no main effect of *persona*, $F(1, 44) = 3.36$, $p = .074$, $\eta_p^2 = .07$, and no interaction effect, $F(2, 88) = 1.96$, $p = .146$, $\eta_p^2 = .04$. Follow-up t-tests, however, did not show any significant differences between conversational tasks (all $p > .05$ after correction for multiple comparisons).

Overall, the extraverted persona was rated as more likable than the introverted persona and agents were rated as most likable during the knowledge test. Closeness was not significantly affected by persona or task.

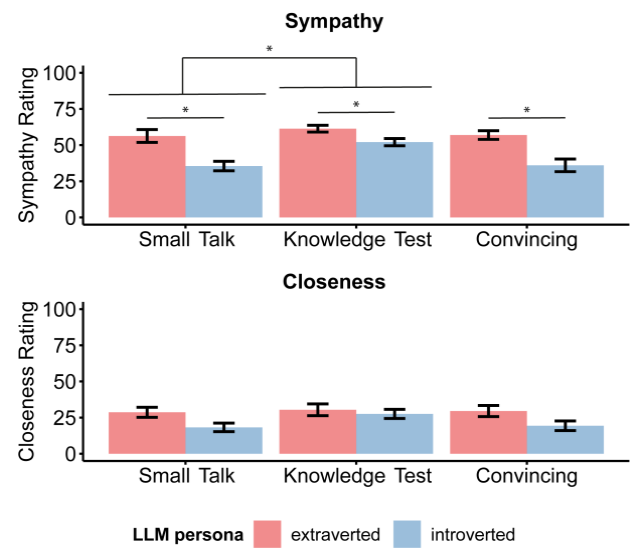

*Figure 2: Social evaluation of the virtual agent in ratings of sympathy (top) and perceived closeness (bottom) as a function of LLM persona and conversational task. Error bars reflect standard error of the mean.*

### 3.3. Emotional Experience

To characterize the influence of LLM persona and conversational task on emotional experience valence and arousal ratings were analyzed (Figure 3).

ANOVA results for valence ratings revealed a main effect of *persona*, $F(1,44) = 6.08$, $p = .018$, $\eta_p^2 = .12$, but no main effect of *conversational task*, $F(2, 88) = 0.34$, $p = .714$ .001, $\eta_p^2 < .01$, and no interaction effect between *persona* and *conversational task*, $F(2, 88) = 0.33$, $p = .720$, $\eta_p^2 < .01$. Participants interacting with the extraverted persona rated their experience as more pleasant (M = 56.97, SD = 13.64) than participants interacting with the introverted persona (M = 45.01, SD = 18.88), but experience was not influenced by conversational task.

Analysis of arousal ratings revealed only a main effect of *conversational task*, $F(2, 88) = 5.49$, $p = .006$, $\eta_p^2 = .11$, but no main effect of *persona*, $F(1, 44) = 1.03$, $p = .316$, $\eta_p^2 = .02$, and no interaction effect, $F(2, 88) = 1.52$, $p = .224$, $\eta_p^2 = .03$. Participants rated both the knowledge test task (M = 44.96, SD = 22.58) and the convincing task (M = 44.71, SD = 20.77) as more arousing than the small talk task (M = 35.24, SD = 20.07), t(45) = 2.76, p = .022 and t(45) = 2.81, p = .022 respectively. There was no difference in arousal ratings between the knowledge test and the convincing task, t(45) = 0.07, p = .041.

LLM persona and conversational tasks had a differential effect on emotional experience, while interacting with a extraverted persona increased pleasantness compared to interacting with an introverted persona, arousal was mainly driven by conversational task, with convincing and knowledge test being more arousing than the small talk task.

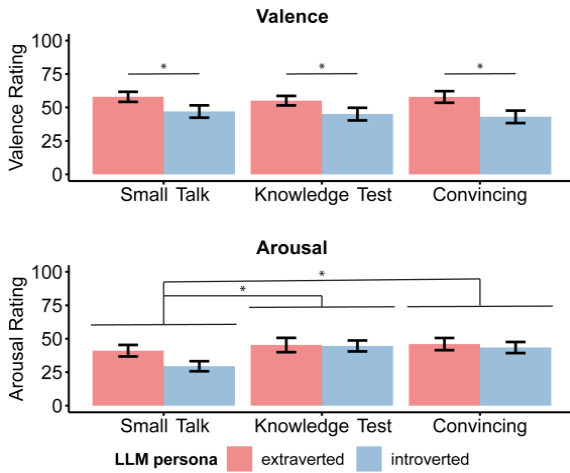

*Figure 3: Participants' emotional experience during social interaction in terms of valence (top) and arousal (bottom) as a function of LLM persona and conversational task. Error bars reflect standard error of the mean.*

### 3.4. Experience of virtual interactions

Another goal was to investigate how LLM persona and conversational task influenced how realistically the interactions with the virtual agent were rated and how social presence was experienced (Figure 4).

ANOVA of the realism ratings revealed a main effect of *conversational task, F(2, 88) = 6.89, p = .002, $\eta_p^2$ = .14*, but no main effect of *persona*, F(1, 44) = 3.43, p = .071, $\eta_p^2$ = .07. However, there was a significant interaction effect of *persona* and *conversational task*, F(2, 88) = 5.07, p = .008, $\eta_p^2$ = .10. Post-hoc Welch t-tests revealed that interaction with the extraverted persona (M = 38.26, SD = 20.14) was rated as more realistic than interaction with the introverted persona (M = 20.52, SD = 17.07) during the convincing task, t(42.85) = 3.22, p = .002, but there was no difference in realism between personas for the small talk (Extraverted: M = 26.91. SD = 16.38; Introverted: 21.78, SD = 16.13), t(43.99) = 1.07, p = .290, and the knowledge test task (Extraverted: M = 35.09, SD = 23.99; Introverted: 32.61, SD = 15.38), t(37.46) = 0.41, p = .679. A follow-up on the main effect of *conversational task* showed that the knowledge test was rated as more realistic than the small talk task, t(45) = 3.38, p = .004, but there were no differences between the small talk and convincing task, t(45) = -1.78, p = .129, and between the knowledge test and convincing task, t(45) = 1.90. p = .129. Overall, persona affected the experienced realism of an interaction only in the convincing task, and interaction during the knowledge test was rated as more realistic than small talk interaction.

In addition, ANOVA results of social presence revealed a main effect of *conversational task*, F(2, 88) = 6.53, p = .002, $\eta_p^2$ = .13, but no main effect of *persona*, F(1, 44) = 1.19, p = .282, $\eta_p^2$ = .03, and no interaction effect, F(2, 88) = 0.14, p = .867, $\eta_p^2$ < .01. Post-hoc t-test revealed that social presence was significantly higher during small talk compared to the knowledge test, t(45) = 3.72, p = .002, but there was no significant difference between small talk and convincing, t(45) = 1.74, p = .137, and between knowledge test and convincing, t(45) = -1.87, p = .137.

The results show that the extraverted compared to introverted LLM persona increased realism ratings specifically for the convincing task. Interestingly, with respect to conversational tasks, the knowledge test was rated as more realistic than small talk but small talk induced a greater feeling of social presence than the knowledge test.

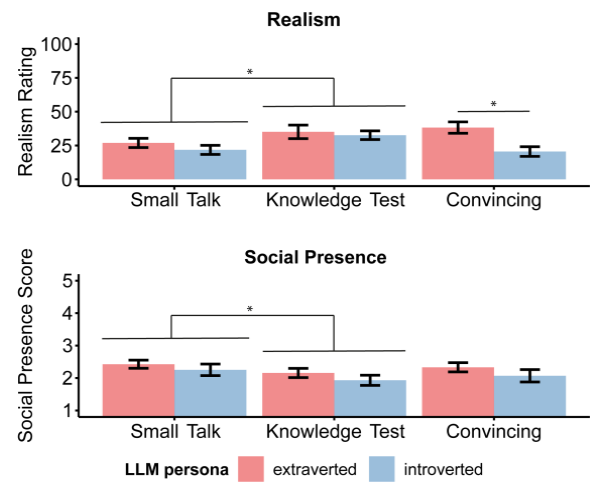

*Figure 4: Evaluation of realism (top) and social presence (bottom) of the virtual interactions as a function of LLM persona and conversational task. Error bars reflect standard error of the mean.*

### 3.5. Behavioral Engagement during Interaction

In addition to self-reports, behavioral parameters were measured to investigate how LLM persona and conversational task affect participants' engagement in the interaction with the virtual agent (Figure 5). Three parameters were assessed for this: (1) The number of pre-defined topics/questions that participants worked through during a conversational task. Note, that a higher number means that participants were faster to switch to the next topic and therefore engaged less with the virtual agent for a given topic. (2) The number of words per topic measured the average number of words that participants spoke during a topic, more words means that participants were more engaged in the conversation. (3) The number of initiated turns per topic measured how many back-and-forth alternations with the virtual agent were initialized by participants for a single topic in average. A higher number of turns indicates more interactive engagement.

First, the number of the pre-defined topics/questions that participants worked through with the virtual agent was analyzed. ANOVA results revealed a significant main effect of *persona*, F(1, 43) = 26.99, p < .001, $\eta_p^2$ = .39, a significant main effect of *conversational task,* F(1.52, 65.36) = 178.08, p < .001, $\eta_p^2$ = .81, but not interaction effect, F(1.52, 65.36) = 0.54, p = .538, $\eta_p^2$ = .01. Participants reached more of the pre-defined topics/questions when interacting with the introverted (M = 14.11, SD = 7.01) compared to the extraverted persona (M = 10.06, SD = 6.34). The highest number of topics/questions was reached in the knowledge test task (M = 19.11, SD = 5.45), followed by the social interaction task (M = 11.24, SD = 4.13), and the convincing task (M = 5.76, SD = 2.89). Post-hoc t-tests showed significant differences between all conversational tasks (ps < .001). Overall, participants reached a higher number of topics/questions when interactions were shorter and when the persona manipulation resulted in shorter responses of the virtual agent (i.e. in the introverted condition).

Next, the average number of words per pre-defined topic/question was analyzed. The ANOVA revealed an effect of *persona, F(1, 43) = 16.43, p < .001,* $\eta_p^2$ *= .28,* and *conversational task*, F(1.26, 54.29) = 168.17, p < .001, $\eta_p^2$ = .80, as well as an interaction between *persona* and *conversational task*, F(1.26, 54.29) = 6.96, p = .007, $\eta_p^2$ = .14. Post-hoc t-tests were conducted to follow-up on the interaction effect. There was a significant effect of persona for each level of conversational task, with greater number of words for the extraverted persona compared to the introverted persona (Small Talk: t(30.84) = 2.92, p = .013; Knowledge Test: t(30.53) = 2.43, p = .021; Convincing: t(42.01) = 3.34, p = .005). However, an analysis of differences between conversational tasks revealed that the extraverted persona group produced significantly more words in the convincing than the knowledge task compared to the introverted persona group, t(42.11) = 3.12, p = .010, but groups did not differ in differences between the small talk and knowledge test task, t(31.67) = 2.14, p = .067, and differences between the small talk and convincing task, t(39.76) = -2.20, p = .068. Overall, participants spoke more words with the extraverted persona agent than the introverted persona agent, and more words in the convincing task than the small talk and knowledge test task. Finally, participants talking with the extraverted persona specifically increased the number of words in the convincing task relative to the knowledge task compared to participants talking to the introverted persona agent.

In a further analysis, the number of turns initiated by the participant per topic/question was investigated. ANOVA results showed a significant main effect of *persona*, F(1,43) = 4.36, p <. 001, $\eta_p^2$ = .09, a significant main effect of conversational task, F(1.39, 59.57) = 111.62, p < .001, $\eta_p^2$ = .72, but there was no significant interaction effect, F(1.39, 59.57) = 1.16, p = .304, $\eta_p^2$ = .03. Participants interacting with the extraverted persona initiated more turns per question (M = 2.78, SD = 2.17) than participants interacting with the introverted persona (M = 2.27, SD = 1.70). With respect to conversational task, participants initiated more turns per question in the convincing task (M = 4.40, SD = 1.95) than in the small talk task (M = 2.46, SD = 1.00), t(44) = 6.66, p < .001, and more turns per question in the small talk compared to knowledge test task (M = 0.72, SD = 0.24), t(44) = 11.85, p < .001. Therefore, participants were more likely continue a conversation about a particular topic when interacting with the extraverted compared to the introverted agent and there the most attempts to continue the conversation were observed in the convincing task followed by the small talk task.

Overall, the behavioral parameters show a consistent pattern, with the extraverted LLM persona resulting in more interactive engagement than the introverted persona. This effect was more pronounced in the convincing task for the number of words but not the number of turns.

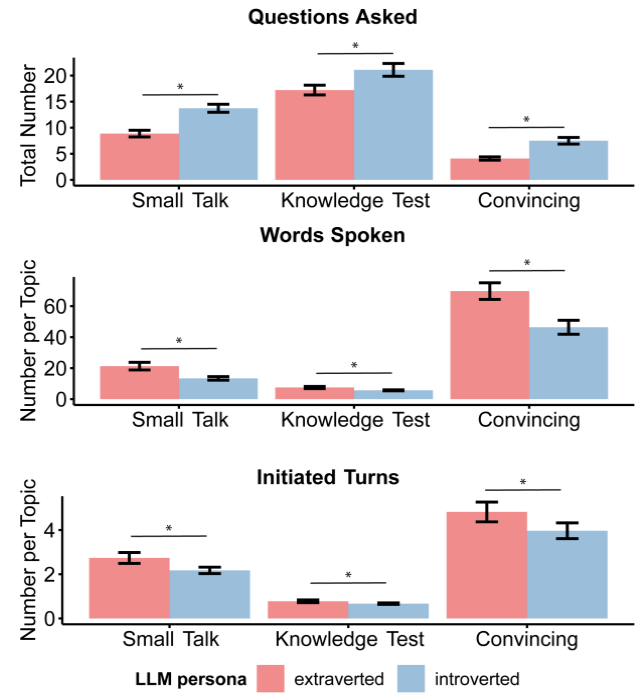

*Figure 5: Behavioral parameters during conversations as a function of LLM persona and conversational task. Error bars reflect standard error of the mean.*

### 3.6. Using AI for help in the knowledge test

During the knowledge test, participants had to answer questions and were free to ask the virtual agent for help. Questions were grouped into easy and diffult questions. The percentage of questions for which participants asked for help was then analyzed with a mixed ANOVA using the between-subject factor *persona* and the within-subject factor *question difficulty*. The analysis revealed a significant main effect of *difficulty*, F(1, 44) = 666.49, p < .001, $\eta_p^2$ = .94, but no effect of *persona*, F(1, 44) = 1.01, p = .321, $\eta_p^2$ = .02, and no interaction effect between *persona* and *difficulty*, F(1, 44) = 0.28, p = .601, $\eta_p^2$ < .01. Therefore participants asked the LLM-controlled agent more frequently for help when difficult questions had to be answered (M = 90.03 %, SD = 9.91 %) compared to when easy questions had to be answered (M = 23.12 %, SD = 17.47 %), but asking for help was not influenced by persona of the LLM controlled virtual agent.

In addition, participants rated how confident they were about the correctness of their answers, either for questions answered with or without help of the virtual agent (factor *help*). Analysis of these confidence ratings revealed a main effect of *help,* F(1, 44) = 11.19, p = .002, $\eta_p^2$ = .20, but no effect of *persona,* F(1, 44) = 0.71, p = .404, $\eta_p^2$ = .02, and no interaction effect, F(1, 44) = 0.11, p = .740, $\eta_p^2$ < .01.

Participants were more confident about the correctness of their answers when they asked the LLM controlled agent for help (M = 82.89, SD = 19.07) compared to when they answered the question without help (M = 70.12, SD = 16.80) but this was not modulated by personality of the virtual agent.

Overall, the data show that participants were more likely to ask the LLM for help when hard compared to easy questions had to be answered and they were more confident in the answers when they had asked for help, however, there was no influence of the LLM persona.

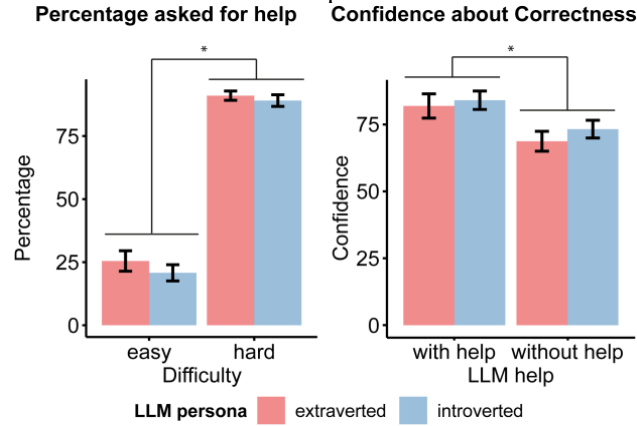

*Figure 6: Behavior and Confidence in the knowledge test conversational task. Graph left shows the percentage of answered questions for which participants asked the virtual agent for help as a function of question difficulty and persona. Graph right shows how participants rated their confidence in the correctness of their answer as a function of whether they had asked the virtual agent for help and persona. All error bars reflect the standard error of the mean.*

## *4.* **Discussion**

The goal of this study was to investigate how the persona of an LLM controlled embodied conversational agent modulates experience and behavior in social interactions across different conversational tasks. Participants interacted with either a extraverted or an introverted LLM persona during a small talk, a knowledge test or a convincing task. In order to characterize the influence of these manipulations we assessed social-evaluative processes, emotional experience, and evaluation of the VR scenario, as well as behavioral parameters such as interactive engagement and participants' tendency to ask the LLM for help. Overall, we found consistent evidence for an effect of persona in line with our hypotheses that conversing with the extraverted persona resulted in a more positive social evaluation, a more positive emotional experience and greater interactive engagement than conversing with the introverted persona. There was, however, no consistent interaction effect between LLM persona and conversational tasks. The effect of LLM persona was only modulated by conversational tasks with respect to the experienced realism and number of words, in a sense that the relative increase in realism and engagement from introverted to extraverted was most pronounced in the convincing task. This finding did not confirm the hypothesis that the effect of extraversion would be greater in the small talk compared to the knowledge test and convincing task. Furthermore, in contrast to our hypothesis the small talk was not more pleasant than the knowledge test and convincing task. Finally, while we found no support for the hypothesis that a extraverted LLM personality would increase participants tendency to ask the virtual agent for help during the knowledge test, we did find that raising question difficulty strongly increased the tendency to ask for help and that participants were more confident in the answers when they had asked the LLM controlled agent for help. Overall, the present study demonstrates that persona of a LLM embodied conversational agent strongly influences experience and behavior in social interactions, although more data are required to characterize the influence of conversational tasks.

Prompting persona traits in a LLM strongly affects social evaluation of the virtual agent, emotional experience, and interactive engagement. Interaction with the extraverted persona increased the feeling of sympathy towards the virtual agent compared to the introverted persona, similarly as in real-word interactions (Wortman & Wood, 2011). In line with findings from chat-based interactions (Liu & Sundar, 2018) our results demonstrate that users are sensitive to the communicative style of a virtual interactive agent and adapt their behavior and socio-emotional processing accordingly. The more positive evaluation of the extraverted persona in comparison to the introverted persona is in line with previous studies which found more positive social evaluations (i.e. liking) in real-life for persons scoring high on sociable aspects of extraversion such as being talkative and joyful (Wortman & Wood, 2011). However, it should be noted that extraversion may have a strong influence on an initial positive evaluation but may have a weaker influence on the formation of social bonds across longer interactions (Harris & Vazire, 2016). Interestingly, this was also reflected in the present study, where we did not observe an influence of LLM persona on interpersonal closeness. However, such deeper social connections may develop over longer periods of time as has been demonstrated for multi-session interactions with a chatbot (Araujo & Bol, 2024). Overall, the present study demonstrates that manipulations of extraversion via LLM prompts impact socio-evaluative processes of face-to-face social interactions in immersive virtual environments and simulate findings from human-human interaction.

A further target of our study was to shed light on the influence of task characteristics on social interaction. Interestingly, participants experienced the argumentation focused conversation as equally pleasant yet more arousing than the small talk conversation. This finding suggests that the exchange of arguments was experienced as a deeper conversation rather than an adverse disputation and can be explained by the fact that the LLM was trained to be a helpful assistant and therefore responses of the model were always respectful and constructive. Furthermore, we observed a modulation of the impact of persona by task for the evaluation of interactive realism. Interaction with the extraverted persona was experienced as more realistic than interaction with the introverted persona in the deeper argumentative conversations, but persona did not affect realism in the other conversational tasks. This suggests that persona manipulations become more important when more elaborative conversations with agents are implemented. In contrast, the effects of persona on social evaluation as well as emotional experience were consistent across conversational tasks suggesting that participants did not change their initial evaluations. Finally, it should be noted that virtual agents were evaluated as most positive during the knowledge test suggesting that receiving help by a conversational agent per se results in a more favorable evaluation of the agent. Overall, these data demonstrate that task demands and role of an agent can affect the experience of interactions with LLM-controlled agents and can also modulate the effects of agent characteristics.

Notably, the experimental manipulations of persona and task affected not only experience of social interactions, but also had a direct impact on participants behavior. Interacting with a extraverted compared to an introverted persona agent resulted in greater interactive engagement. While the underlying mechanisms remain to be explored, it is possible that participants tendency to reciprocate the agents' communicative style may drive the effect (Frisch & Giulianelli, 2024; Kroczek & Gunter, 2020; Schoot et al., 2016). The present study also supports findings from previous studies, in which manipulation of LLM personality resulted in behavioral outcomes (Lim et al., 2024). This is relevant for many applications, e.g. in education or health care, which aim at activating users. Incorporating LLM-controlled embodied conversational agents may also be promising for VR applications that provide mental health interventions (Bell et al., 2024; Graf et al., 2024; Herbener et al., 2024). Besides interactive engagement across conversational tasks, the study also investigated participants tendency to ask the conversational agent for help to answer knowledge questions. Using LLMs to acquire knowledge is a typical use task for human-AI interaction. Our findings did not reveal an effect of persona, neither for easy nor difficult questions. Interestingly, however, participants had greater confidence in their answer when they relied on the conversational agent (cf. Wester et al., 2024). This demonstrates that there is a general tendency to accept the answers provided by a LLM as true (Zhou et al., 2019). Overall, the present study provided evidence that interacting with an virtually embodied and LLM-controlled conversational agent directly impacts users' behavior and this effect can be further modulated by manipulating the agent's characteristics and communicative style.

Social interactions with LLM-controlled embodied virtual agents are a novel and promising way to study social behavior and human-computer interactions. However, there are some limitations that need to be discussed. First, the conversational tasks were presented in a fixed order which may have resulted in carry-over effects as an impression of the agent acquired in the small talk task might have persisted across the other tasks (Digirolamo & Hintzman, 1997). With this limitation in mind, it should be noted that the order of the present study was set to resemble a natural interaction between humans, where a short phase of getting to know each other typically precedes deeper interaction. Furthermore, we found that some parameters differed only in later conversational tasks, suggesting that task-specific effects were preserved. Another limitation is that the virtual agents did not produce non-verbal behavior or emotional expressions together with the speech output. Non-verbal behaviors have been demonstrated to have a crucial role for communication and social interactive behavior (Kroczek et al., 2024; Kroczek & Mühlberger, 2023), future studies should therefore aim at incorporating both non-verbal and verbal behavior resulting in a more naturalistic social interactions (for an example see Llanes-Jurado et al., 2024). The present study highlights the use of LLMs to control conversations with embodied virtual agents and demonstrates that an agent's persona characteristics and communicative style can have a profound impact on experience and behavior across different conversational tasks that mimics real-life. This has implications for the use of embodied conversational agents in applications related to mental health and for research focusing on social behavior.